\documentclass[usegraphicx,usenatbib,useapjfonts,apj,twocolappendix]{emulateapj}

\def\cmtres{\mbox{cm$^{-3}$}}
\def\esf{\mbox{{$\epsilon_{\rm SF}$}}}
\def\erg{\mbox{erg}}

\def\FbU{\mbox{$F_{U}$}}
\def\fg{\mbox{$f_{g}$}}
\def\Fgh{\mbox{$F_{g,h}$}}

\def\kev{\mbox{keV}} 
\def\Ke{\mbox{K}}

\def\lcdm{\mbox{$\Lambda$CDM}}
\def\lwdm{\mbox{$\Lambda$WDM}}

\def\mh{\mbox{$M_{\rm v}$}}
 
\def\mpch{\mbox{$h^{-1}$Mpc}} 
\def\ms{\mbox{$M_{s}$}}
 
\def\msunh{\mbox{$h^{-1}$M$_\odot$}} 
\def\msun{\mbox{M$_\odot$}}

\def\nsf{\mbox{$n_{\rm SF}$}}
\def\ome{\mbox{$\Omega_m$}} 
\def\omeb{\mbox{$\Omega_b$}}
\def\omel{\mbox{$\Omega_\Lambda$}}

\def\re{\mbox{$R_{e}$}}
\def\rh{\mbox{$R_{\rm v}$}}

\def\rhosf{\mbox{$\rho_{\rm SF}$}} 

\def\sige{\mbox{$\sigma_8$}} 
\def\Tsf{\mbox{{$T_{\rm SF}$}}}
\def\vmax{\mbox{$V_{\rm max}$}} 

\def\vth{\mbox{$v_{\rm th}$}} 
\def\mwdm{\mbox{$m_{\rm WDM}$}}
\def\sigDM{\mbox{$\sigma_{\rm SI}$}}
\def\lcut{\mbox{$\lambda_{\rm fs}$}}
\def\Mhm{\mbox{$M_{f}$}}
\def\wdma{\mbox{WDM$_{1.2}$}}
\def\wdmb{\mbox{WDM$_{3.0}$}}

\def\mg{\mbox{$M_{g}$}}
\def\zh{\mbox{$z_{f,h}$}}
\def\mb{\mbox{$M_{b}$}}

\newcommand{\jcap}{JCAP}

\def\mathnew{\mathsurround=0pt} 
\def\simov#1#2{\lower .5pt\vbox{\baselineskip0pt 
    \lineskip-.5pt\ialign{$\mathnew#1\hfil##\hfil$\crcr#2\crcr\sim\crcr}}}   
\def\simgreat{\mathrel{\mathpalette\simov >}}  
  
\def\'#1{\ifx#1i{\accent"13\i}\else{\accent"13#1}\fi}

\defcitealias{alej2014}{G+2014}

\def\plotancho#1{\includegraphics[width=18cm]{#1}}
\shorttitle{Galaxies formed in WDM halos at the filtering scale}
\shortauthors{Colin et al.}

\begin{document}

\title{Simulations of galaxies formed in warm dark matter halos of masses at the filtering scale}
\author{P. Col\'{\i}n\altaffilmark{1},  V. Avila-Reese\altaffilmark{2}, A. Gonz\'alez-Samaniego\altaffilmark{2}, H. Vel\'azquez\altaffilmark{3}
}

\altaffiltext{1}{Centro de Radioastronom\'{\i}a y Astrof\'{\i}sica, Universidad Nacional 
Aut\'onoma de M\'exico, A.P. 72-3 (Xangari), Morelia, Michoac\'an 58089, M\'exico }

\altaffiltext{2}{Instituto de Astronom\'{\i}a, Universidad Nacional Aut\'onoma de M\'exico, 
A.P. 70-264, 04510, M\'exico, D.F., M\'exico}

\altaffiltext{3}{Instituto de Astronom\'{\i}a, Universidad Nacional Aut\'onoma de M\'exico, 
Apdo. Postal 877, Ensenada, BC, CP 22830, M\'exico}

\begin{abstract}
We present zoom-in N-body + Hydrodynamic simulations of dwarf central galaxies
formed in Warm Dark Matter (WDM) halos  with masses at present-day of $2-4\times 10^{10}$ \msun.
Two different cases are considered, the first one when halo masses are close to the corresponding half-mode 
filtering scale, \Mhm\ (\mwdm =1.2 keV) and the second when
they are 20 to 30 times the corresponding \Mhm\ (\mwdm = 3.0 keV).
The WDM simulations are compared with the respective Cold Dark Matter (CDM) simulations.
The dwarfs formed in halos of masses $(20-30)\Mhm$
have roughly similar properties and evolution than their CDM counterparts; on the
contrary, those formed in halos of masses around \Mhm, are systematically different from
their CDM counterparts. As compared to the CDM dwarfs, they assemble the dark and stellar 
masses later, having mass-weighted stellar ages 1.4--4.8 Gyr younger; their circular
velocity profiles are shallower, with maximal velocities 20--60\% lower;
their stellar distributions are much less centrally concentrated and with
larger effective radii, by factors 1.3--3. The WDM dwarfs at the filtering scale  (\mwdm =1.2 keV) 
have disk-like structures, and end in most cases with higher gas fractions and 
lower stellar-to-total mass ratios than their CDM counterparts. The late halo assembly,
low halo concentrations, and the absence of satellites of the former with respect to the latter,
are at the basis of the differences.  
\end{abstract}
                                                                                                                 
\keywords{cosmology:dark matter --- galaxies:dwarfs --- galaxies:formation --- 
methods:N-body simulations --- methods: Hydrodynamics}

%==========================
\section{Introduction}
\label{intro}
%==========================

The $\Lambda$ cold dark matter (\lcdm) cosmology provides the most accepted
background for studying the process of cosmic structure formation in the Universe. 
The predictions of the \lcdm-based scenario of structure formation are fully consistent 
with observations of the large-scale structure of the present and past Universe, 
including the anisotropies in the cosmic microwave background radiation 
\citep[see][for a review]{Frenk+2012}. However, doubts have been cast on whether 
observations of matter distribution at small --dwarf galactic and subgalactic-- scales are 
consistent with the predictions of the \lcdm\ scenario (see for recent reviews 
e.g., \citealp{Weinberg+2013}; \citealp{delPopolo+2014}). Currently, it is matter of great 
debate whether the potential problems are real or consequence of observational biases 
and/or still poorly understood astrophysical processes at small scales. If they are confirmed, 
introducing variations to the \lcdm\ cosmology will appear as a feasible solution.

From the point of view of initial conditions for the cosmic structure formation,
\lcdm\ is the simplest model. For the \lcdm\ model:
(1) the cut-off scale in the linear mass power spectrum of perturbations due to free streaming, \lcut, is 
many orders of magnitude 
smaller than the resolution reached by current numerical cosmological simulations of galactic halos
so that in practice $\lcut=0$ and  hierarchical cosmic structure formation proceeds at all scales;
(2) the relic thermal velocities of the CDM particles, \vth, are very small, 
so in practice \vth=0 is assumed;
(3) since CDM particles are non-baryonic, they do not interact 
electromagnetically, and are assumed to have a negligible self-interaction cross section, $\sigDM= 0$,
constituting the CDM in practice a pure collisionless fluid; and 
(4) the statistical distribution of the primordial overdensity perturbations is assumed to be
Gaussian. 
Therefore, the relaxation of any of the above listed assumptions implies 
necessarily the introduction of free parameters
in the initial conditions of cosmic structure formation as would be
\lcut, \vth, \sigDM, or the skewness and kurtosis in the primordial density perturbations distribution.

More than a decade ago, high-resolution N-body cosmological simulations were performed 
to explore how substructure, inner density profiles and shapes of halos were affected
when one or several of the \lcdm\ assumptions listed above were relaxed. Specifically, in these simulations 
were introduced: (a) a cut-off in the power spectrum and/or non-negligible thermal velocities 
in the dark particles (in concordance with the $\Lambda$ warm dark matter, \lwdm, cosmology; 
\citealp{Colin+2000}; \citealp{Avila-Reese+2001}; 
see also, \citealp{Bode+2001}, \citealp{Knebe+2002});
(b) a non-negligible self-interaction with constant and velocity-dependent particle cross sections 
\citep[][see also Spergel \& Steinhardt 2000; Firmani et al. 2000]{Yoshida+2000,Colin+2002}; (c) and 
non-Gaussian initial perturbations, positively or negatively skewed 
\citep{Avila-Reese+2003}.

Among alternative cosmologies, the most popular is the \lwdm\ one with a power spectrum filtered
at scales corresponding to dwarf galaxies. As N-body simulations show, in this case the amount of
substructure
in Milky Way-sized halos is considerably reduced \citep[e.g.,][]{Colin+2000,Bode+2001,Knebe+2002,
Maccio+2010,Kennedy+2014}, the abundance of low-circular velocity halos hosting dwarf galaxies 
is lowered \citep[e.g.,][]{Zavala+2009,Papastergis+2011}, 
and, although the halos/subhalos do not present shallow cores at the scales of astrophysical interest, 
they are less concentrated and with lower maximum circular velocities than their CDM counterparts
\citep[][]{Avila-Reese+2001,Colin+2008,Lovell+2012,Schneider+2012,Anderhalden+2013}.
These and other effects make the \lwdm\ model an appealing alternative for alleviating the potential 
problems of the \lcdm\ model at small scales, while conserving its successes at larger scales. 

The main constraint to the \lwdm\ scenario comes from the comparison of the results of
WDM hydrodynamic simulations in the quasi-linear regime with the Ly-$\alpha$ flux power spectrum 
of high-redshift quasars \citep[][]{Narayanan+2000,Viel+2005}, though
these comparisons are not free of uncertainties and limitations \citep[see e.g.,][]{deVega+2014}.
Depending on the nature of the WDM particle, thermal, sterile neutrino, etc., a lower limit to its mass, 
\mwdm, can be established from the Ly-$\alpha$ forest analysis, which implies an upper limit to the 
damping scale in the mass power spectrum. Several updated estimates were presented
recently in the literature \citep[e.g.,][]{Viel+2013}. Based on the constrains of the latter authors 
(for a thermal relic particle, \mwdm\ should be $\gtrsim 3.3$ keV at the 2$\sigma$ level), 
\citet{Schneider+2014} conclude that the upper limit in the damping is at so small scales that the 
allowed \lwdm\ models would not be already able to solve the potential problems of \lcdm. 

So far, most studies on galaxy properties in the \lwdm\ scenario were based on dark-matter-only 
simulations or a combination of these kind of simulations with semi-analytic models 
\citep[for the latter see e.g.,][]{Maccio+2010,Menci+2012,Kang+2013}. However,  by their own 
nature, these approaches can not take into account the effects of the non-linear baryonic physics 
on the evolution and dynamics of the halos, which can be important. Thus, inferences based on 
the analysis of dark-matter-only simulations (and, perhaps, semi-analytic models) are necessarily 
limited when comparing with observations \citep[see e.g.][]{Kang+2013}. It is then important to 
go beyond those techniques and perform full N-body + Hydrodynamics simulations. An interesting 
question than one can ask is how much differ the evolution and properties of galaxies formed in 
the \lwdm\ scenario from those formed in the \lcdm\ one.  At this point, it is important to recognize 
that the dark-matter structure evolution is expected to be very similar in both scenarios at scales 
much larger than the filtering one, with differences appearing gradually at scales approaching this scale. 

The so-called half-mode wavelength or its corresponding mass, \Mhm, is commonly chosen as the relevant {\it filtering}
scale at which WDM halo abundance and properties start to significantly deviate from the CDM case (see for the exact 
definition and references Section \ref{cosmodel}). In this paper, we present a set of zoom-in N-body + 
Hydrodynamics simulations of (dwarf) galaxies formed in WDM scenario in halos with masses 
$\lesssim\Mhm$ and $20-30 \Mhm$, and compare them with their CDM counterparts. Recently, 
\citet[][see also Libeskind et al. 2013]{Herpich+2014} reported WDM simulations of this kind but for 
three halos of masses significantly larger than the filtering masses corresponding to their WDM particle masses
(\mwdm= 1, 2, and 5 keV). This is very likely the reason why the evolution of their WDM galaxies did not  
differ significantly from the CDM counterparts. After completion of our study, it appeared a preprint by 
\citet{Governato+2014}, where the authors present a simulation of {\it one} dwarf galaxy formed in a 
$\sim 10^{10}$ \msun\ halo, both in WDM and CDM cosmologies. For the WDM cosmology, \mwdm=2 keV
 was used, which implies that their system is $\sim 2$ times larger than the filtering mass. 

Here, our goal is to explore the evolution and properties of dwarf galaxies formed in {\it halos of masses similar to} 
the filtering mass \Mhm\  that corresponds to a thermal WDM particle mass of 1.2 keV. 
These are expected to be among the most abundant halos in  this WDM scenario; below $\sim 0.5\Mhm$, the 
halo mass function strongly decreases, and the structures, rather than virialized halos, are isolated 3D enhancements 
not assembled hierarchically \citep[][]{Angulo+2013}. 
 The evolution and properties of the galaxy-halo systems around the filtering mass {\it might in several aspects
be generic regardless of the value of this mass}; though, astrophysical processes 
such as stellar feedback and reionization could make the extrapolation of our results to
other filtering or thermal WDM particle masses inadequate.

In Section 2, the cosmological background and the used WDM models are presented. 
The details of the code and simulations performed here are given in Section 3. 
The properties and evolution of the simulated WDM dwarf galaxies and their corresponding CDM
ones are presented in Section 4. A summary of the results and further discussion are presented in Section 5.

%============================================
\section{The cosmological models}
\label{cosmodel}
%============================================

The cosmological background used in our numerical simulations
is a flat, low-density model with $\ome = 0.3$, $\omeb=0.045$, $\omel = 0.7$,  and $h = 0.7$.
For the CDM initial power spectrum, $P(k)_{CDM}$, we 
adopt the approximation by \citet{KH97}, which was obtained as a direct fit 
of the power spectrum calculated using a Boltzmann code. For the scales
studied in this paper, and even larger ones, this approximation is very accurate. 
In the case of WDM, the power spectrum at large scales is essentially that of the 
CDM, but at small scales the power is systematically reduced due to the free-streaming 
damping. The transfer function $T^2_{WDM} (k)$ describes such a deviation from the
CDM power spectrum, 
\begin{equation}
P_{WDM} (k) = T^2_{WDM} (k) P_{CDM} (k).
\label{P-S}
\end{equation}
The CDM or WDM power spectra are normalized to
$\sige = 0.8$, a value close to that estimated from the {\it Planck} mission \citep{PAR2013};
 \sige\ is the rms of $z=0$ mass perturbations estimated with the top-hat window of radius $8 \mpch$. 

The free-streaming of collisionless particles erase dark matter perturbations
below a scale given by the properties of the dark matter particle. Here, we will refer
to the case of fully thermalized particles at decoupling  as {\it thermal relics}.  
A simple analysis gives an estimate of the comoving
length at which thermal particles diffuse out \citep[e.g.,][]{KolbTurner, Schneider+2012}: 
\begin{equation}
\lambda_{fs} \simeq 0.4\left( \frac{\mwdm}{\kev} \right)^{-4/3} 
\left( \frac{\Omega_{WDM} h^2}{0.135} \right)^{1/3} [\msunh].
\label{eq:lamfs}
\end{equation}
However, in order to calculate the whole processed power spectrum, 
the coupled Boltzmann relativistic system of equations for the various species of matter and radiation
should be numerically solved. Here, we adopt the WDM transfer function
given in \citet{Viel+2005}:
\begin{equation}
T_{WDM} (k) = \left[ 1 + (\alpha k )^{2.0\nu} \right]^{-5.0/\nu},
\label{TF}
\end{equation} 
where $\nu = 1.12$ and the parameter $\alpha$ is related to $\mwdm$, $\Omega_{WDM}$, 
and $h$ through 
\begin{equation}  
\alpha = a \left( \frac{\mwdm}{1 \kev} \right)^b \left( \frac{\Omega_{WDM}}{0.25} \right)^c
\left( \frac{h}{0.7} \right)^d \mpch, \label{eq:alfa}
\end{equation} 
with $a= 0.049$, $b=-1.11$, $c=0.11$, $d=1.22$. 
In eq. (\ref{TF}), $\alpha$ is a characteristic scale length that can be related to 
an effective free-streaming scale, $\lambda_{\rm fs}^{\rm eff}\equiv \alpha$ \citep[e.g.,][]{Schneider+2012}.
The corresponding effective free-streaming mass is then:
\begin{equation}
M_{\rm fs} = \frac{4 \pi}{3} \bar\rho \left(\frac{\lambda_{\rm fs}^{\rm eff}}{2} \right)^3,
\label{eq:Mfs}
\end{equation}
where $\bar\rho$ is the present-day background density. 
The primordial density perturbations below $M_{\rm fs}$ are expected to be completely erased, 
while perturbations with masses up to thousand times $M_{\rm fs}$ can be significantly 
affected by the damping process. It is common, on the other hand, to define
a characteristic scale below which the linear WDM power spectrum start to deviate significantly
from the CDM one \citep{SomDol01,Avila-Reese+2001}.  
Following \citet[][]{Avila-Reese+2001}, we define the half-mode wavenumber, $k_{\rm hm}$, for which 
$T^2_{WDM}(k)= 0.5$; i.e., where the value of the power spectrum of the
WDM model is half that of the corresponding CDM one.
The associated half-mode {\it filtering} mass is given by:
\begin{equation}
\Mhm = \frac{4 \pi}{3} \bar\rho \left(\frac{\lambda_{\rm hm}}{2} \right)^3, 
\label{Mhm}
\end{equation}
where $\lambda_{\rm hm}=2\pi/k_{\rm hm}$ is the comoving half-mode length.\footnote{Note that some 
authors define the half-mode wavenumber as $T_{WDM}(k)= 0.5$ \citep[e.g.,][]{Schneider+2012}, 
which implies a smaller \Mhm\ than in our case.} This filtering mass scale, which is much larger than
$M_{\rm fs}$, is where one expects the abundance and properties of the halos to start to significantly 
deviate from the CDM case \citep[][]{Colin+2008,Smith+2011,Menci+2012,Schneider+2012,Benson+2013,Angulo+2013}.
At masses around \Mhm, the abundance of halos already falls below, by a factor of $\sim 2$, 
that of the corresponding CDM one, reaching its relatively shallow peak at $\sim 0.5$ \Mhm. 
Thus, the most abundant halos in the WDM cosmogony are those 
of masses around to \Mhm. 

Structures of masses close to \Mhm\ can be unambiguously defined
as approximately spherical virialized objects that resemble those seen in CDM simulations,
albeit with some differences; for instance, they are less concentrated  \citep[e.g.,][]{Avila-Reese+2001,Angulo+2013}. 
Moreover, halos of masses $\gg \Mhm$ are expected to assemble hierarchically, sharing the
same properties as their CDM counterparts.  Systems of masses several times smaller than \Mhm\ but
larger than $M_{\rm fs}$ can be defined as "protohalos", that is halos that are not fully formed, 
but show clear isolated
3D enhancements \citep{Angulo+2013}. At masses close to $M_{\rm fs}$, these authors report structures that appear
as clear failures of their halo finder algorithm, these include outer caustics of large halos 
and dense sheets and filaments, where the collapse of a further axis has just started.

%==========================
\begin{figure}
\vspace{10.4cm}
\includegraphics{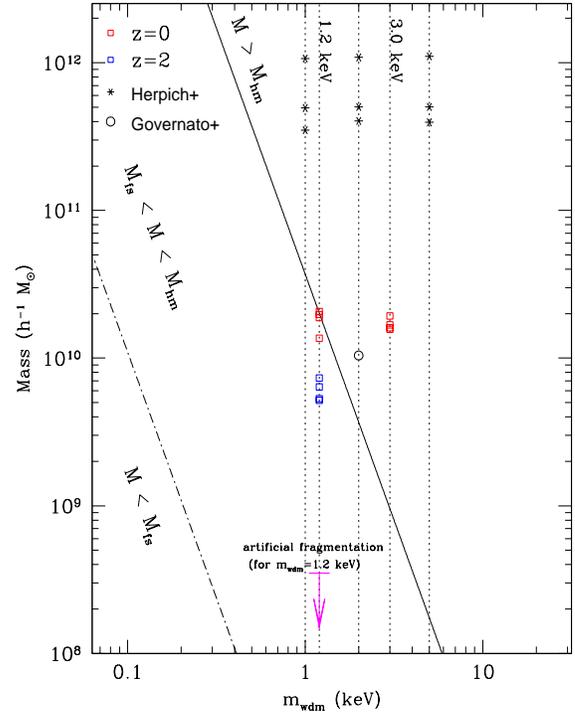}
\caption{The free-streaming and half-mode (filtering) mass scales as a function of the thermal relic
particle mass, \mwdm\ (for a sterile neutrino particle, see eq. \ref{steril-niu}). The squares
show where our simulations lie at $z=0$ (red) and at $z=2$ (blue; only for the 1.2 keV case). Even
at $z=2$ our simulations are far from the artificial fragmentation scale for a filtering corresponding
to 1.2 keV (downwards arrow). The starred symbols and the open circle correspond to 
the simulations presented in \citet{Herpich+2014} and \citet{Governato+2014} at $z=0$.  }
\label{fig1} 
\end{figure}
%==========================

Following \citet[][]{Schneider+2014}, in Fig. \ref{fig1}, we show the dependence
of \Mhm\ and $M_{\rm fs}$ on the thermal WDM particle mass \mwdm. 
For a given \mwdm\ and at any epoch, objects located well above the
solid line ($M = \Mhm$) are halos assembled hierarchically, as in the CDM cosmogony. 
Objects of masses close to \Mhm, on the other hand, can be considered ``normal'' halos. 
However, their early assembly is already affected by the filtering at small 
scales; for example, they form later than their CDM counterparts. 
The objects that \citet{Angulo+2013} define as ``protohalos'' start to apear as the mass 
decreases below \Mhm\ and dominate the population when $M \ll \Mhm$. 
In these ``protohalos'', the hierarchical assembly fails notoriously
\citep[e.g.,][]{Schneider+2012}. Finally, below the dot-dashed line ($M=M_{\rm fs}$), no 
structure formation is expected. However, in N-body  simulations it is possible to find
structures smaller than $M_{\rm fs}$ but they are actually artificial
\citep[see e.g.,][]{Avila-Reese+2001, Bode+2001,gotz03,Knebe+2003,WW2007,Schneider+2013,Angulo+2013}.

In Fig. \ref{fig1} we plot our eight simulated systems (presented below) with red squares as well 
as those simulated by \citet{Herpich+2014} and \citet{Governato+2014}, starred and open
circle symbols (their M$_{200}$ masses were multiplied
by 1.24 so as to take into account our different definition of virial mass). These masses are at $z=0$;
we also plot the virial masses of our simulated systems at $z=2$ (blue squares). The  downwards arrow
indicates the mass scale where spurious structures would form in our simulations with \mwdm=1.2 keV 
(for \mwdm=3 keV, it is at a smaller mass) due to numerical fragmentation, according to the criterion 
given by \citet{WW2007}. This criterion depends on the filtering scale and the resolution of 
the numerical simulation. As can be seen, our halos lie above  the downwards arrow even at $z=2$. 
Yet, it might be that numerical artifacts are present at the highest redshifts, 
when the progenitor masses are very small.

The relations show in Fig. \ref{fig1} are for thermal relic particles. Popular candidates for WDM 
are also the sterile and right-handed neutrinos, particles proposed to never been in thermal equilibrium 
(\citealp{Dodelson+1994}; \citealp{Shi+1999}; \citealp{Abazajian+2001}; see \citealp{Boyarsky+2009} for a review). 
\citet{Viel+2005} provided a relation between the  non-resonantly produced sterile neutrino mass and the mass
of the thermal relic particle such that the transfer function for this kind of sterile neutrino
can be calculated according to eq. (\ref{TF}). The relation is given by
\begin{equation}
m_{\nu_s}= 4.43\ \kev (\mwdm/1\ \kev)^{4/3} (\Omega_{WDM} h^2)^{-1/3}.
\label{steril-niu}
\end{equation}
Thus, for a given cosmological background, we can use Fig. \ref{fig1} 
also for the case of the  non-resonantly produced sterile neutrino, after going from \mwdm\ to $m_{\nu_s}$.  
Another particle candidate for WDM is the gravitino 
\citep[e.g.,][]{Pagels+1982,Ellis+1984,Moroi+1993}. Weakly interacting massive particles (WIMPs, the 
most favored candidates for CDM), if produced in non-thermal processes, can also have large free-streaming
lengths and emulate the power spectrum of WDM \citep[e.g.,][]{Lin+2001,He+2013}. 

Finally, thermal WDM particles are expected to have a relic velocity dispersion, 
which could affect the inner structure of halos due to the Liouville theorem limit for the phase
density \citep[][]{Hogan+2000}. However, this velocity dispersion is very small for 
reasonable mass candidates ($\sim 3.0$ keV) and even for thermal particle masses as low 
as $\mwdm\sim 1$ keV it does not affect significantly the inner structure of WDM halos
\citep{Avila-Reese+2001,Colin+2008,Maccio+2012}. Therefore, we assume a zero thermal 
velocity dispersion, \vth=0, in our simulations.

%=====================
\section{The code and the simulations}
%=====================

We have carried out a set of N-body + Hydrodynamics simulations of low-mass halos
using the zoom-in technique in both the \lwdm\ cosmology and its counterpart
the \lcdm\ one. 
The simulations were run using the Adaptive Refinement Tree (ART)
N-body/hydrodynamic code \citep[][]{KKK97,Kravtsov03}. 
The code includes gas cooling, star formation (SF), 
stellar feedback, advection of metals, and a UV heating background source.
The cooling and heating rates incorporate Compton heating/cooling, atomic and molecular 
cooling, UV heating from a cosmological background radiation \citep{HM96}, and are
tabulated for a temperature range of $10^2 < T < 10^9\ \Ke$ and a grid of densities,
metallicities, and redshifts using the CLOUDY code \citep[version 96b4]{Ferland98}. 

The SF and feedback processes (subgrid physics) are implemented in the
code as discussed in detail in \citet{Colin+2010} and \citet{Avila-Reese+2011}. For
completeness, they are briefly summarized below. The SF takes place in those cells for which  
$T < \Tsf$ and $\rho_g > \rhosf$, where $T$ and $\rho_g$ are the temperature and 
density of the gas, respectively, and $\Tsf$ and $\rhosf$ are the temperature and
density threshold, respectively. Here, we use the same values of the 
$\Tsf$ and $\nsf$ parameters as in \citet{Avila-Reese+2011}; namely, 9000 \Ke\ and 6 \cmtres,
respectively, where \nsf\ is the density threshold in hydrogen atoms per
cubic centimeter; see \citet{Avila-Reese+2011}, as well as \citet{Colin+2010} and 
\citet{Gonzalez+2014} (hereafter G+2014), for a discussion on the choice of these values, 
in particular \nsf.  A stellar particle of mass $m_* = \esf\ m_g$ is placed in a grid cell every time 
the above conditions are simultaneously satisfied, where $m_g$ is the gas mass 
in the cell and \esf\ is a parameter that measures the local efficiency
by which gas is converted into stars. As in \citet{Avila-Reese+2011}, we set $\esf = 0.5$.

We use the ``explosive'' stellar thermal feedback recipe, according to which each star more 
massive than 8 \msun\ injects instantaneously into the cell, 
where the particle is located, $E_{{\rm SN+Wind}} = 2 \times 10^{51}\ \erg$ 
of {\it thermal} energy; half of this energy is assumed to come from the type-II SN 
and half from the shocked stellar winds. This energy provided by the stellar feedback
raises the temperature of the cell to values $\simgreat 10^7\ \Ke$; the
precise value depends on the assumed initial mass function (IMF),
the amount of energy assumed to be dumped by each massive star, and 
the value of the \esf\ parameter. On the other hand, each $8 \msun$ ejects 
$1.3 \msun$ of metals. For the assumed \citet{MS79} IMF, a stellar particle 
of $10^5\ \msun$ produces 749 type-II SNe.

In our previous works, we have delayed the radiative cooling for some time
(typically between 10 and 40 Myr) in those cells where young stellar particles are, 
in order to avoid overcooling due to, for instance, resolution limitations. However, at
the current resolution reached by our simulations and for 
the typical densities found in the SF cells ($\sim 10\ \cmtres$), which in
turn depend on \nsf, the cooling time is actually much larger 
than the crossing time \citep{DS2012}. Thus, for the 
simulations used in this study, 
switching-off the cooling temporarily is expected to
have only a minor effect on the properties
of the simulated galaxies. We have carried out some tests and 
verified that this is the case. However, we decided to keep 
in the code this cooling delay when running the \lwdm\ simulations 
because some of the corresponding CDM galaxies, to 
be compared with the WDM ones, were run with this prescription
(G+2014).  

\begin{table*}
 \begin{center}
  \caption{Physical properties of WDM runs at z = 0}
  \begin{tabular}{@{}cccccccccccc@{}}
  \hline
  Name   & log(\mh) &  log(\ms)\footnote{Mass within 0.1\rh (the same applies for \mg);}  & log(\mg)   & \vmax  & 
\re\footnote{Radius that encloses half of the stellar mass within 0.1\rh;}  & \rh  & \fg\footnote{$\fg\equiv \mg/(\mg+\ms)$;}  &  
$M_{g, cold}/\mg$\footnote{The amount of cold gas inside the galaxy in units of \mg} & 
D/T\footnote{Ratio of the mass contained in the high-angular momentum disk stars with respect to the total stellar mass;}  
&\zh\footnote{Redshift at which the given halo acquired one third of its present-day mass.}  & T\footnote{Stellar mass-weighted average age.}  \\
& (\msun) &(\msun)  &(\msun)   &(km $s^{-1}$)  &(kpc)  &(kpc) &  &  &  &  &(Gyr) \\  %($10^{-3} \msun$ $\rm yr^{-1}$) \\
  \hline  
   \multicolumn{12}{c}{ $m_{p} = 0.0$ keV (CDM)} \\
  \hline
   Dw3    & 10.46  & 8.73   & 8.07  & 56.07  & 1.02  & 78.50 & 0.18 & 0.81 & 0.01  & 2.30  & 9.67 \\ %0.34  \\ %Dw8
   Dw4    & 10.38  & 8.38   & 8.61  & 52.87  & 0.80  & 73.64 & 0.63 & 0.67 & 0.19  & 1.90  & 8.90 \\ %1.08  \\  %Dw7
   Dw5    & 10.46  & 8.55   & 9.08  & 61.52  & 1.10  & 77.34 & 0.77 & 0.82 & 0.49  & 1.90  & 8.79 \\ %1.69  \\ %Dw9
   Dw7    & 10.39  & 8.21   & 8.60  & 43.44  & 2.20  & 73.90 & 0.71 & 0.75 & 0.59  & 1.70  & 6.79 \\ %7.76  \\  %Dw11
   Dwn1    & 10.63  & 8.90   & 9.23  & 63.71  & 2.83  & 89.77 & 0.68 & 0.91 & 0.66  & 2.10  & 6.84 \\ %41.7  \\ %Dwn01
   Dwn2    & 10.58  & 8.72   & 9.13  & 62.73  & 1.25  & 84.20 & 0.72 & 0.86 & 0.57  & 1.50  & 9.45 \\ %0.01  \\ %Dwn02
  \hline
   \multicolumn{12}{c}{ $m_{p} = 1.2$ keV (\wdma)} \\
  \hline
   Dw3    & 10.43  & 8.74   & 8.92  & 47.87  & 2.03  & 75.24 & 0.60 & 0.86 & 0.43 & 1.85  & 4.89 \\ %4.11  \\ %Dw8
   Dw5    & 10.29  & 7.80   & 8.28  & 37.57  & 1.51  & 68.86 & 0.76 & 0.56 & 0.46 & 1.86  & 6.42 \\ %0.01  \\ %Dw9
   Dwn1   & 10.47  & 8.43   & 8.29  & 43.23  & 3.61  & 78.79 & 0.42 & 0.42 & 0.65 & 1.85  & 5.48 \\ %0.01  \\ %Dwn01
   Dwn2   & 10.45  & 8.41   & 9.00  & 49.34  & 3.85  & 76.60 & 0.80 & 0.80 & 0.73 & 1.50  & 6.26 \\ % 0.01  \\ %Dwn02
\hline
  \multicolumn{12}{c}{ $m_{p} = 3.0$ keV (\wdmb)} \\
\hline
   Dw3    & 10.36  & 8.67   & 6.27  & 50.83  & 0.86   & 72.54 & 0.004& 0.00 & 0.00  & 2.60  & 9.69 \\ %0.01  \\ %Dw8
   Dw4    & 10.35  & 8.21   & 7.75  & 48.07  & 0.80  & 72.50 & 0.26 & 0.20 & 0.14 & 2.10  & 10.09 \\ %1.64  \\ %Dw7
   Dw5    & 10.44  & 8.47   & 9.07  & 58.23  & 1.44  & 76.23 & 0.80 & 0.87 & 0.55 & 2.10  & 7.95 \\ %2.85  \\ %Dw9
   Dw7    & 10.38  & 7.99   & 8.73  & 42.80  & 3.08  & 73.23 & 0.85 & 0.60 & 0.78 & 1.60  & 5.69 \\ %0.01  \\ %Dw11
  \hline
 \end{tabular}
 \label{table}
\end{center}
\end{table*}

%=================================
\subsection{The zoom-in simulations}
\label{simulations}

The aim of this paper is to explore the evolution of galaxies 
formed in WDM halos of masses around the half-mode (filtering) scale \Mhm. As discussed
in Section \ref{cosmodel},  \Mhm\ %(see eq. \ref{Mhm})
is a characteristic scale, where the properties and abundance of the WDM halos start to depart 
significantly from those of their CDM counterparts. 

Here, we will study one particular value of \Mhm, corresponding to dwarf-galaxy scales. 
At a qualitative level, the results might apply to other values of \Mhm.

According to eqs. (\ref{Mhm}) and (\ref{TF}), a half-mode mass of $2\times 10^{10}$ \msunh\
is associated to a thermal relic particle of $\mwdm=1.2$ keV (see Fig. \ref{fig1}).
In G+2014, a set of seven zoom-in \lcdm\ N-body+ Hydrodynamics simulations that end 
up with around this mass were analyzed with the purpose of exploring the effects of different
halo mass assembly histories (MAHs) on the evolution and properties of central dwarf galaxies. 
These low-mass simulated galaxies have enough resolution (see below) so
that an analysis of their internal structural properties can be done with
some confidence. Therefore, in order to study systems at the half-mode mass scale and
compare them with the CDM results,  we carry out here zoom-in simulations of some
of the G+2014 runs but using the WDM power spectrum corresponding to 
$\mwdm=1.2$ keV (case \wdma\ hereafter).   We also carry out simulations of the 
same systems for a WDM power spectrum corresponding to 
$\mwdm=3$ keV (case \wdmb\ hereafter), in order to explore whether the
properties of simulated galaxies in halos much more massive than \Mhm\
tend to be similar to those of galaxies formed in CDM halos. For this case,
\Mhm= $9.6\times 10^8$ \msunh\ and thus, the simulated systems
are about 20-30 times \Mhm.

The \lwdm\ simulations performed here have the same random seed and box size 
as the \lcdm\ simulations in G+2014. Therefore, all the target WDM halos/galaxies
here have their CDM counterpart simulations. The box used in G+2014 has $L_{box}=10\ \mpch$ 
per side and a root grid of $128^3$ cells. We first set the multiple-mass species initial conditions 
with the code PMstartM \citep{KKBP01} and
then run a low--mass resolution simulation with the N-body ART code 
\citep{KKK97}. A spherical region of radius three times the virial 
radius \rh\ of the selected halo is chosen. The virial radius is defined as the radius that encloses a mean 
density equal to $\Delta_{vir}$ times the mean density of the universe, where $\Delta_{vir}$ is 
obtained from the spherical top-hat collapse model. The Lagrangian region corresponding to the $z = 0$ spherical
volume is identified at $z= 50$ and resampled with additional small-scale waves
\citep{KKBP01}. The new zoom-in simulation is then run with the hydrodynamic/N-body
version of ART.  The number of DM
particles in the high-resolution zone changes from halo to halo but it is
between 500 thousand and one million. The mass
per particle $m_p$ in the highest resolution region is $6.6 \times 10^4\ \msunh$ and increases
for the DM N-body only runs to $7.8  \times 10^4\ \msunh$. 

In ART, the grid is refined recursively as the matter distribution evolves. 
The runs use a DM or gas density criteria to refine. 
In the CDM runs presented in G+2014, the cell is refined 
when its mass in DM exceeds $1.3\ m_p$ or the mass in gas is 
higher than 1.4$\FbU m_p$, where $\FbU\equiv \omeb/\ome$
is the universal baryon fraction. For the WDM runs, we have decided to
use a less aggressive refinement (it acts as a softening of very small structures)
in an attempt to eliminate artificial fragments\footnote{Although our target 
halos/galaxies have masses at $z\sim 0$ much larger than
the scale where fake structures would form (see Fig. \ref{fig1}),
the artificial fragments could affect our results at very high redshifts.}, which are 
known to arise due to the finite number of particles and the resolved cut-off 
of the power spectrum (see above for references). 
Thus, we refine cells only until they reach a mass eight times the previous value of
$1.3\ m_p$ in DM or 1.4$\FbU m_p$ in gas. 
To make sure that the less aggressive refinement does not introduce 
significant differences in the simulations, we resimulated
some of the CDM runs with the less aggressive refinement
setting. A comparison between 
the CDM halos/galaxies obtained with the aggressive and soft
refinements was done for some of the dwarfs and roughly the same evolution 
and properties were obtained. In the Appendix, we show and discuss the case for run Dw3.

%===========================
\begin{figure*}
\plotancho{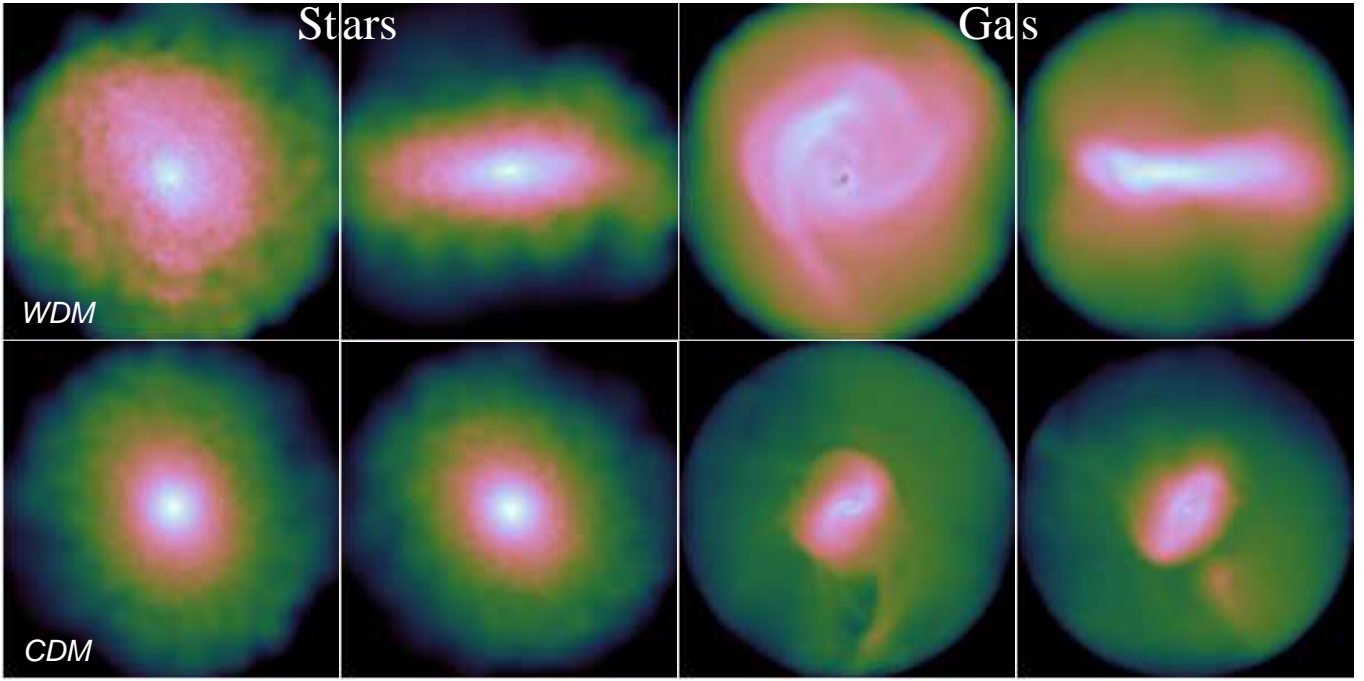}
\epsscale{1.0}
\caption[fig:fig2]{Stellar (two leftmost panels) and gas (two rightmost panels) projected distributions
at $z=0$ for the dwarf Dw3 in the \wdma\ (top) and CDM (bottom) runs within a sphere of radius 0.15 \rh.
The first and third (second and fourth) columns are projections in the face-on (edge-on) planes 
of the galaxy. The \wdma\ run clearly show a more extended gaseous and stellar structure
than its CDM counterpart.
\label{fig2}}
\end{figure*}
%===========================

As in G+2014, in the hydrodynamic simulations presented in 
this paper, the root grid of $128^3$ cubic cells is immediately refined unconditionally to the 
third level, corresponding to an effective 
grid size of $1024^3$.  Although we formally set the maximum refinement level to 11,
which implies a minimum cell size of 55 comoving pc, this is not reached in practice 
in the WDM runs with $\mwdm=1.2$ \kev. In general, the number of cells inside \rh\
depends on the mass of the halo/galaxy, cosmology, and on the kind of refinement 
that was used, but it is roughly about one million and reduces by a factor of eight or 
so for the less aggressive refinement.

As mentioned above,  our study is focused on dwarfs formed in WDM halos that at $z=0$ 
have masses close to \Mhm\ and much larger than \Mhm, and on comparing them with 
their CDM counterparts.
From G+2014 we have selected the CDM systems named there as Dw3, Dw4, 
Dw5 and Dw7, each one with a different halo MAH but with about the same 
present-day virial masses, $\mh =1.5-3\times 10^{10}$ \msunh.
Unfortunately, the halo finder\footnote{We use a variant of the bound density 
maxima (BDM) halo finder algorithm of \citet{Klypin+1999}, kindly provided by 
A. Kravtsov, and run it on the dark matter particles in order to identify the 
dark matter halos or subhalos. The central galaxy is then
centered at the position of the corresponding most massive halo.}
could not identify the halos corresponding to dwarfs Dw4 and Dw7 
in the WDM simulation with $\mwdm=1.2$ \kev. Hence,
only the systems Dw3 and Dw5 have been simulated in all cosmologies: CDM and WDM with
$\mwdm=1.2$ and 3 \kev. The  systems Dw4 and Dw7 were run in WDM with 
$\mwdm=3$ \kev, in which case the corresponding
\Mhm\ is much smaller than the masses of the simulated objects. 
In order to have more WDM systems of masses around \Mhm, we have
identified in the 1.2 keV WDM box two more distinct halos, around these 
masses, and performed the corresponding zoom-in hydrodynamical simulations 
(dwarfs Dwn1 and Dwn2). The CDM simulations for these systems with the aggressive 
refinement were also run for comparison. 

In Table 1, we present all the runs studied in this paper and summarize their main 
present-day properties.  The WDM runs are shown in Fig. \ref{fig1}. 
As far as we are aware of, our simulations are the only ones, 
perform within the full N-body + hydrodynamical scheme,  that focus on
halo/galaxy masses that at present-day are close to the filtering scale \Mhm\ 
(corresponding specifically to \mwdm=1.2 keV).
The galaxy properties (\ms, stellar galaxy half-mass radius \re, SFR, etc.) are computed within a sphere of 
$0.1$\rh\ radius. We notice that the \re values reported in Table 1 of G+2014 
were erroneously boosted by a factor 1/$h$; the values presented in Table 1 here are the correct ones. 
This radius 0.1 \rh\ contains essentially all the stars and cold gas of the simulated central galaxy. 
The contamination of satellites or other substructures at this radius is negligible. 
On the other hand, because the outer stellar mass density profiles decrease strongly 
with radius in most of the runs, the galaxy stellar mass would not differ significantly had we
measure it at ``aperture'' radii slightly smaller than 0.1\rh\ by, for example, 20-50\%.
The disk-to-total ratio (D/T) is found using a kinematic decomposition of the stellar galaxy
into an spheroid and a disk. The mass of the spheroid, $M_{sph}$, is defined as two times 
the mass of the stellar particles inside 0.1\rh\ that have negative
spin values (counter-rotate), it implicitly assumes that the spin distribution 
of the spheroid is symmetric around zero. The D/T is then defined as $(M_s - M_{sph})/M_s$.

%=========================================================
\section{Results} 
\label{results}
%=========================================================

\subsection{General properties}

All of our zoom-in simulations are for {\it distinct} halos that at the present epoch end up with 
virial masses of $\approx 1.5-3\times 10^{10}$ \msunh; the dwarf
galaxies inside these halos are therefore {\it centrals}.
Those halos of masses around \Mhm\ (runs \wdma)
are devoid of substructures and have mass distributions less concentrated
than their CDM and \wdmb\ counterparts.
In Fig. \ref{fig2}, we plot the 2D stellar and gaseous distributions at $z=0$ for dwarf Dw3 
in the \wdma\ (top) and CDM (bottom) runs. Projections in the face-on (first and third 
columns) and edge-on (second and fourth columns) planes of the galaxy are shown.
The FITS images of these projections were obtained with 
TIPSY\footnote{http://www-hpcc.astro.washington.edu/tools/tipsy/tipsy.html}. 
In these images, the galaxy disk lies on the plane perpendicular to
the angular momentum vector of the gas cells that are within a sphere of radius 0.15\rh,
centered on the center of mass of the stellar particles.  
We use then the  DS9 visualization program\footnote{http://ds9.si.edu/site/Home.html} 
to create the images by color coding the density of the respective components within a fixed 
range of values for a fair comparison between the panels.

From a visual inspection of Fig. \ref{fig2}, we see that
the \wdma\ dwarf galaxy has a more extended and less centrally concentrated
stellar mass distribution than in the CDM case. Indeed, the \re\ of the former is 2 times
larger than the one of the latter (see Table 1). Moreover, the former has a more disk-like
structure than the latter. Indeed, we measure a disk-to-total mass ratio, D/T, 
of 0.43 vs 0.01 (Table 1). A more pronounced disk-like structure for the former than for the latter
is also seen in the gas distribution. The \wdma\ dwarf at the half-mode mass is gas rich (\fg=0.60) and 
it has an extended and a low-surface density gaseous disk, unlike the CDM case, where 
the dwarf is gas poor (\fg=0.18) and has a compact gas
distribution. The spatial temperature distribution
of the gas is also quite different: in the CDM case the gas in the galaxy is colder 
than the one in the dwarf at the half-mode mass scale, while the gas in the corona
is hotter.

According to Table 1, the present-day dwarfs Dw3, Dw5, Dwn1 and Dwn2,
formed in halos of masses around \Mhm, are systematically more extended (larger \re),
have a lower stellar mass and maximum circular velocities, \vmax,
and, in most of cases, end up with a higher gas fractions  %and D/T ratios 
than their CDM counterparts.  These differences are likely a consequence of 
{\it the later assembly, lower concentrations and absence of mergers of the halos 
at the filtering mass}.  However,  due to the complex and non-linear subgrid 
physics, small variations in the non-linear evolutionary processes can also 
produce large differences and shifts in the galaxy properties at any given epoch. 
This is why we have simulated several systems to verify that the differences 
between  the \wdma\ dwarfs and their CDM counterparts
are systematical and not due to small variations in a particular case. Moreover, 
although the systems with masses much larger than \Mhm\ (runs \wdmb) show some
differences with respect to their CDM counterparts, these
are already small and do not follow a systematical trend as in the 
case of the \wdma\ runs. For example, in some \wdmb\ runs 
\fg, \re\ or D/T are larger in the \wdmb\ runs than in their CDM counterparts, while in others 
they are smaller.  This very likely means that the differences are due to small variations
in the non-linear evolution rather than due to the (small) differences in the initial power spectrum.
Yet, \vmax\ is systematically lower in all \wdmb\ runs, but not by much as seen in the case 
of the \wdma\ ones. {\it The depth of the gravitational potential of the systems seems 
to be the main property systematically affected by the filtering in the power 
spectrum of fluctuations.}

%=========================================================

\begin{figure*}
\vspace{11.3cm}
\includegraphics{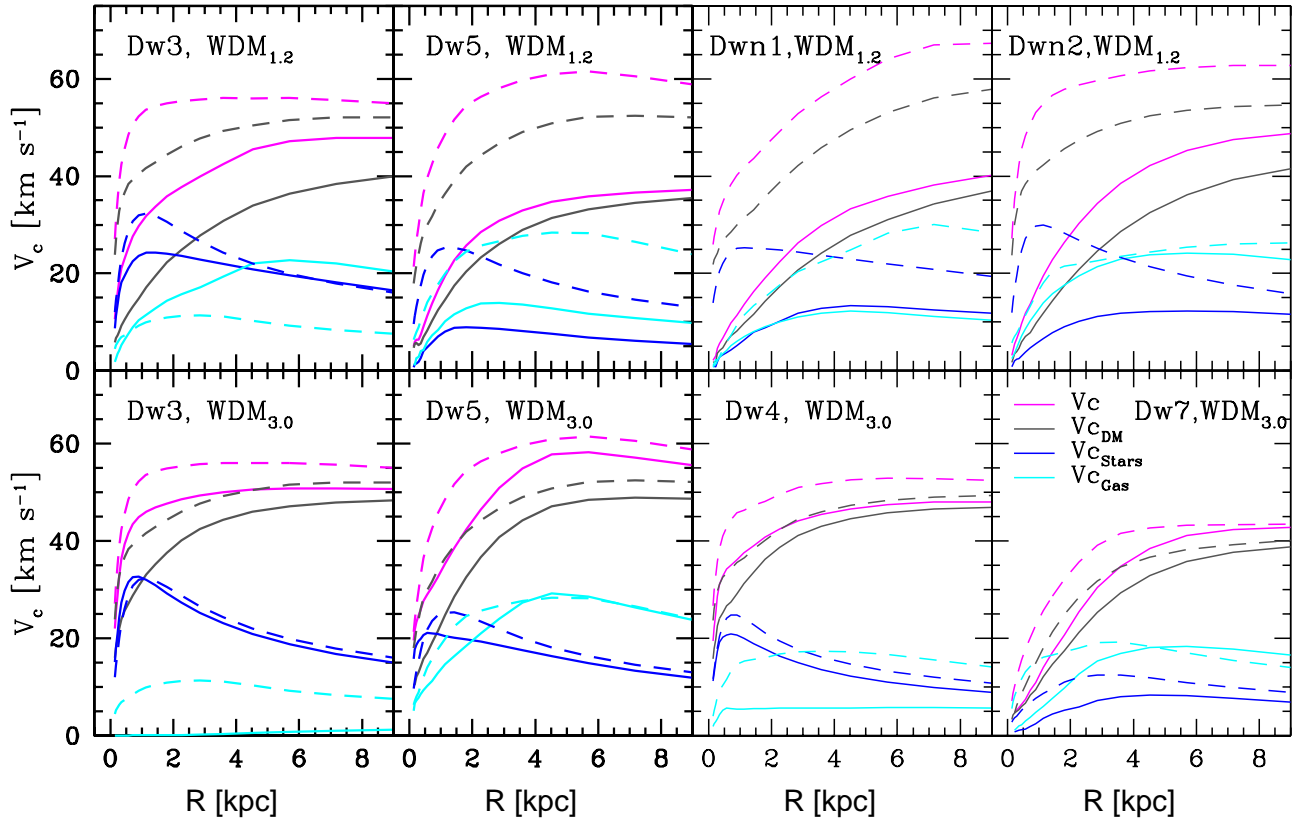}
\caption[fig:fig3]{Circular velocity profiles and their decomposition for the WDM runs
(solid lines) and their CDM counterparts (dashed lines) at $z=0$.  In the upper panels are
plotted the profiles for 
the systems of masses around the filtering scale \Mhm\ and in the lower panels for
those systems of masses 20 to 30 times \Mhm. We denote with magenta, gray, blue, and cyan lines 
the total, halo, stellar, and gaseous circular velocities, respectively. }
\label{fig3} 
\end{figure*}

%=========================================================

%============================
\subsubsection{Radial distributions}

In this subsection, we explore in more detail the present-day inner structure and dynamics of the simulated dwarfs.
Solid lines in Fig. \ref{fig3} show the total circular velocity profiles, $V_c(r)$, and their decompositions into 
DM, stars and gas (gray, blue and cyan lines, respectively) for the systems of mass around
\Mhm\ (runs \wdma, upper panels) and of mass ($20-30)\times\Mhm$ (runs \wdmb, lower panels).
The corresponding CDM dwarfs are also shown with dashed lines using the
same color coding.
The total circular velocity profiles (magenta lines) for  dwarfs formed in halos at the 
filtering scale are {\it shallower in the 
center than the CDM counterparts with lower values of \vmax\ by} 20-60\%. 
These differences can be seen even for dwarfs formed in the \wdmb\ cosmogony, though
they are small, showing that systems formed in halos much larger than \Mhm\
tend to be similar to those formed in the CDM scenario. 
%=========================================================

\begin{figure*}
\vspace{11.0cm}
\includegraphics{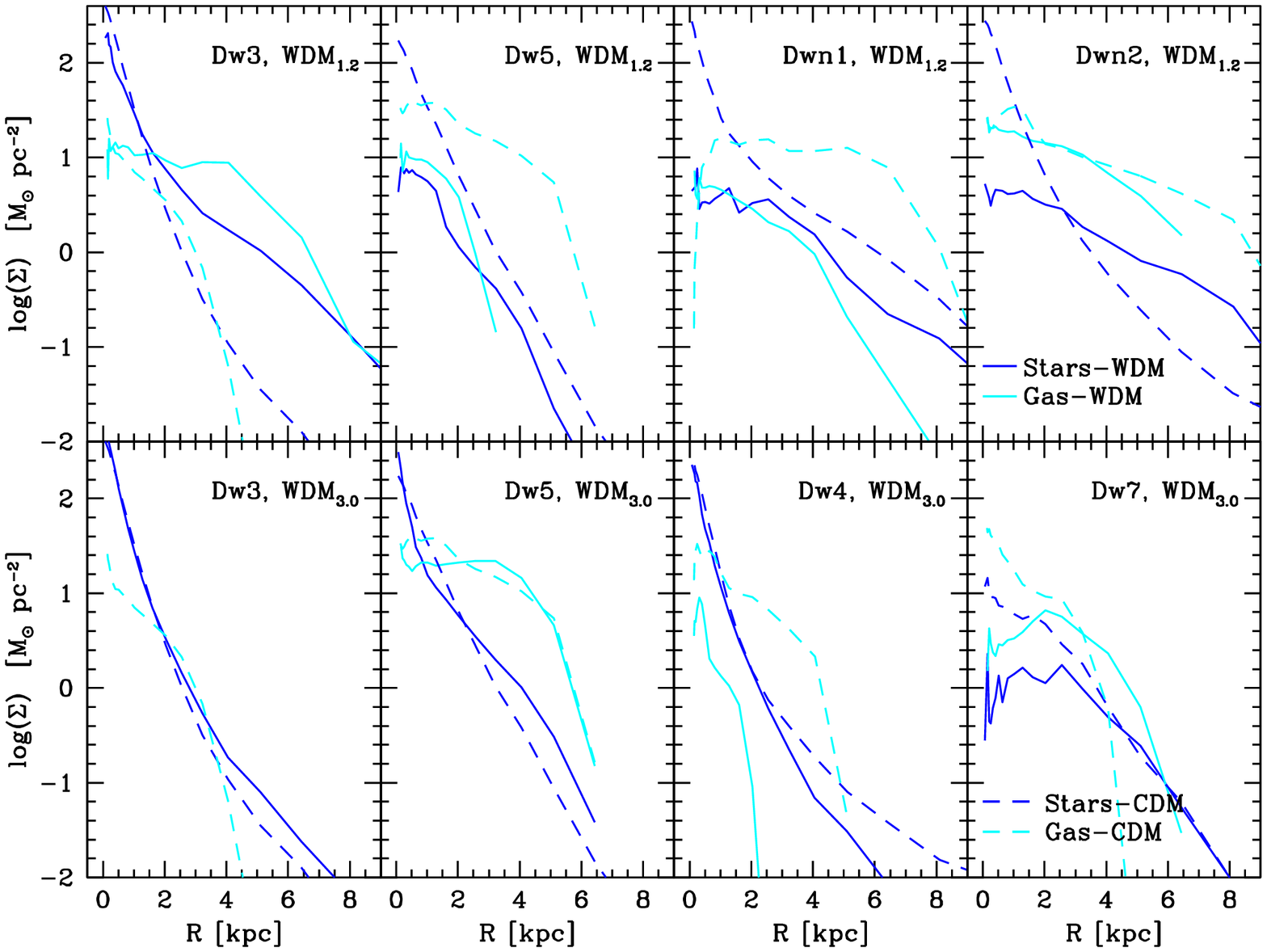}
\caption[fig:fig4]{Stellar (blue lines) and gas (cyan lines) surface density profiles of the 
WDM dwarfs (solid lines), compared to their CDM counterparts (dashed lines).
The upper and lower panels are for the \wdma\ and \wdmb\ runs, respectively. 
}
\label{fig4} 
\end{figure*}

%=========================================================

The stellar $V_c(r)$ components (blue lines) of the \wdma\ runs are significantly lower and less peaked 
than the CDM cases. In all simulations, the halo component dominates,
at least from radii larger than $\sim 0.1-0.7 $ \re. The circular velocities of the gaseous component 
(cyan lines) tend also to be less peaked in the \wdma\ runs than in the CDM ones. However, while 
in the CDM case their inner contributions to the total circular velocity lie below
that of the stellar component, except for dwarf Dw7, 
in the case of the \wdma\ runs, the $V_c(r)$ of the gaseous component 
is similar or dominates at all radii over the stellar one, 
except in the inner region of dwarf Dw3  (\wdma\ dwarfs are more gaseous than the analogous CDM dwarfs,
see Table 1). In any case, the baryonic (stellar + gaseous) contribution 
to the total $V_c(r)$ is more important and {\it more centrally concentrated} in all CDM runs than in the
\wdma\ ones. Therefore, the circular velocities of the WDM systems of masses around \Mhm\
are shallower than the CDM ones mainly because {\it less centrally concentrated baryonic galaxies form
in the former simulations} (see also Fig. \ref{fig4} below). 

On the other hand, \vmax\ is significantly lower in the \wdma\ runs than in the CDM ones mainly 
because the corresponding pure DM halos are already less concentrated in the WDM case than in the CDM one 
\citep[e.g,][]{Avila-Reese+2001,Lovell+2012}.  However, the astrophysical processes in both
cases are also expected to produce different effects on the inner dynamics of the galaxy-halo systems,
which could increase/reduce the differences in \vmax, as well as in the innermost dynamics of the galaxy-halo
systems (for instance, the formation or not of shallow cores). We will study in detail this question 
elsewhere. 

In Fig. \ref{fig4}, the stellar (blue) and cold gas (cyan) surface density (SD) profiles of the \wdma\ 
 and \wdmb\ runs (solid lines) are compared to those of their CDM counterparts (dashed lines). 
The most noticeable difference in the stellar SD profiles between the \wdma\ runs and the
corresponding CDM ones  (upper panels) is that {\it the inner regions of the former are significantly lower}.  
While the CDM dwarfs have a central peaked stellar density, reminiscent of a bulge-like structure, 
in the \wdma\ ones a flattened SD is seen, except in run Dw3; though, even in this case, 
the CDM dwarf has a more peaked SD (see also Fig. \ref{fig3}).  Regarding the gas SD profiles, 
for the CDM dwarfs, they tend to be more extended than the stellar ones and of lower SDs 
in the center, while for the \wdma\ dwarfs, the gas SD profiles roughly follow the stellar ones, 
except for run Dw3. The CDM dwarfs have higher baryonic (stars+gas) SDs in the center than 
the dwarfs formed in halos at the filtering scale.
 In the case of the \wdmb\ dwarfs (lower panels), their stellar and gas SD profiles tend to be
similar to those of their CDM counterparts; though, for Dw4 and Dw7 the gas SD profiles show
significant differences.

The systematical differences in the stellar SD profiles between the \wdma\ and CDM runs, 
specially in the inner regions, could be the result of many effects. 
One of them might be the angular momentum of the halos in which
galaxies form. We have measured the halo (dark matter particles only) spin parameter\footnote{
The spin parameter is defined as $\lambda = \frac{J|E|^{1/2}}{GM^{5/2}}$, where
$J$, $M$ and $E$ are the total angular mometum, mass, and energy, respectively. This latter
quantity is computed, assuming that the halo is virialized, as -$K$, 
where $K$ is the kinetic energy.}
 $\lambda$ for all the runs at $z=0$ with the following results: runs Dw3 and Dw5 have a lower
spin parameter in the CDM runs than in the \wdma\ ones (0.015 and 0.007 vs. 0.033 and 0.012, 
respectively) but for runs Dwn1 and Dwn2 we found the contrary, CDM runs have a 
higher spin parameter (0.072 and 0.083 versus 0.032 and 0.037, respectively). The last two runs,
specially Dwn2, have a relatively late merger in the CDM simulations that could affect the 
spin parameter, although these mergers happened more than $\sim 4$ Gyr ago. 
We have also measured the spin parameter by using the alternative definition of $\lambda$
introduced in \citet{Bullock+2001}. Although the values of $\lambda$ are different, the trend is the same:
runs Dwn1 and Dwn2 have larger values in the CDM simulations, and Dw3 and Dw5 have smaller values. 
In summary, it is not clear that the halo spin parameter is the reason why the stellar distribution 
is less concentrated in the WDM runs than in the CDM ones. We plan to study in detail elsewhere 
(Avila-Reese et al in prep) the question of the spin parameter in WDM and CDM, both in 
DM-only and hydrodynamical simulations. 

The mechanism responsible for the stellar-SD profile differences between the \wdma\ 
and CDM dwarfs can be traced probably 
to the merging history of the central galaxies \citep[][]{Herpich+2014}. 
The satellite interactions/mergers in the case of CDM simulations drive gas to 
the center, where SF proceeds efficiently, producing a cuspy stellar structure. 
The stellar specific angular momentum is also more likely to decrease in the galaxies 
that suffered mergers since the angular momentum of these merging galaxies 
can cancel each other \citep{Cloet-Osselaer+2014}. These processes do
not happen in the \wdma\ dwarfs because they are practically devoid of satellites.
Besides, the galaxy assembly starts later than in the CDM halos, being the disks 
more gaseous and less susceptible to secular evolutionary processes. 

In summary, dwarfs formed in halos of masses around \Mhm\  in our \wdma\ simulations
have a quite different dynamical history and structural inner properties when compared to 
their CDM counterparts, whereas dwarfs formed in a WDM cosmology but in halos with a mass
much larger than \Mhm, tend to be similar to the corresponding CDM dwarfs. 
The latter is in agreement with the simulation results of \citet{Herpich+2014}.

%===============================
\begin{figure*}
\vspace{10.5cm}
\includegraphics{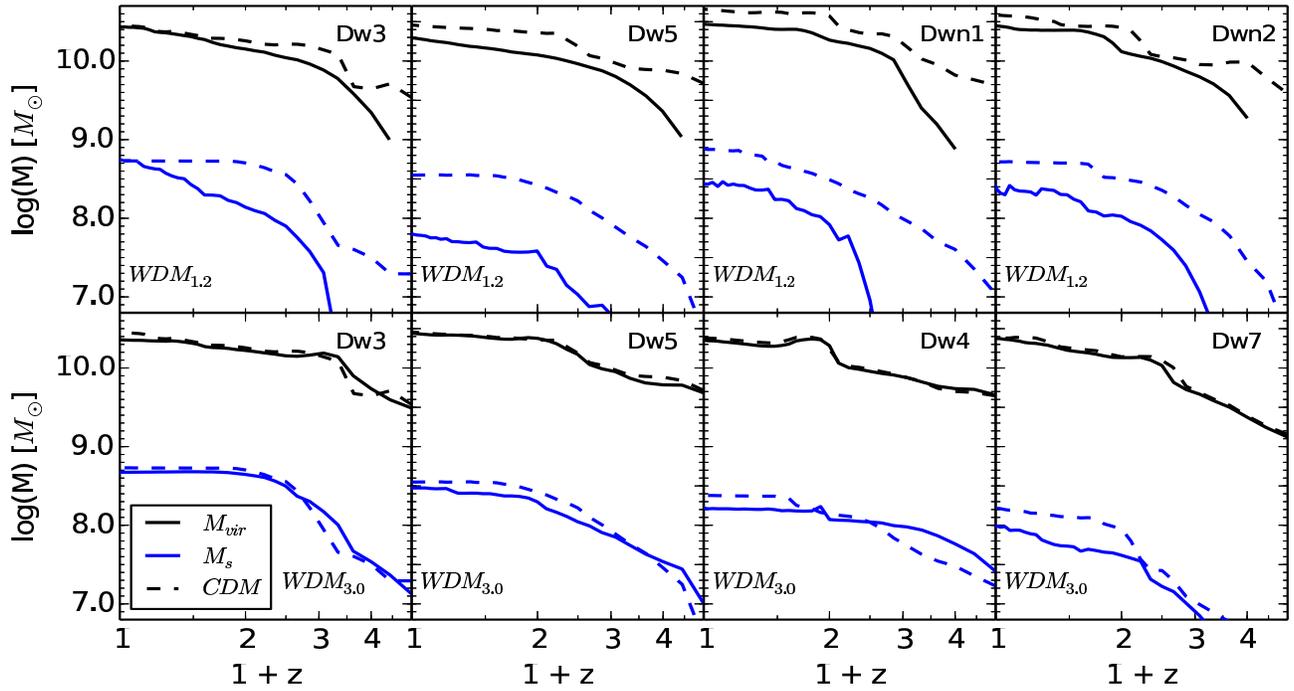}
\caption[fig:fig5]{Mass assembly histories for the simulated dwarf galaxies. In the upper panels 
we show the MAHs for the \wdma\ runs, while in the lower panels are the MAHs of the \wdmb\ runs 
(solid lines). In each case,
the MAHs of the corresponding CDM runs are also plotted (dashed lines). We denote with
black lines the total (virial) MAHs and with blue lines the galaxy stellar MAHs. The
MAHs of the dwarfs formed in halos around \Mhm\ (\wdma\ runs) differ significantly from 
their CDM counterparts.}
\label{fig5} 
\end{figure*}

%=========================================================
\subsection{Mass assembly histories}

In Fig. \ref{fig5} we plot the virial MAHs of the \wdma\ (upper panels) and \wdmb\ (lower
panels) runs along with their corresponding CDM counterparts (black solid and dashed
lines, respectively). As can be seen, the halos around the filtering mass \Mhm\ start to assemble
later and end up with masses slightly smaller than the corresponding CDM ones. 
However, afterwards the former grow faster, specially the runs Dw3 and Dw5. 
Thus, the epoch at which half or one-third of the present-day virial mass is acquired 
is not very different between both cosmologies (see Table 1).
The WDM halos of scales ($20-30)\times \Mhm$ assemble their masses practically in the same 
way as they do in the CDM scenario. 
%========================
\begin{figure*}
\vspace{10.8cm}
\includegraphics{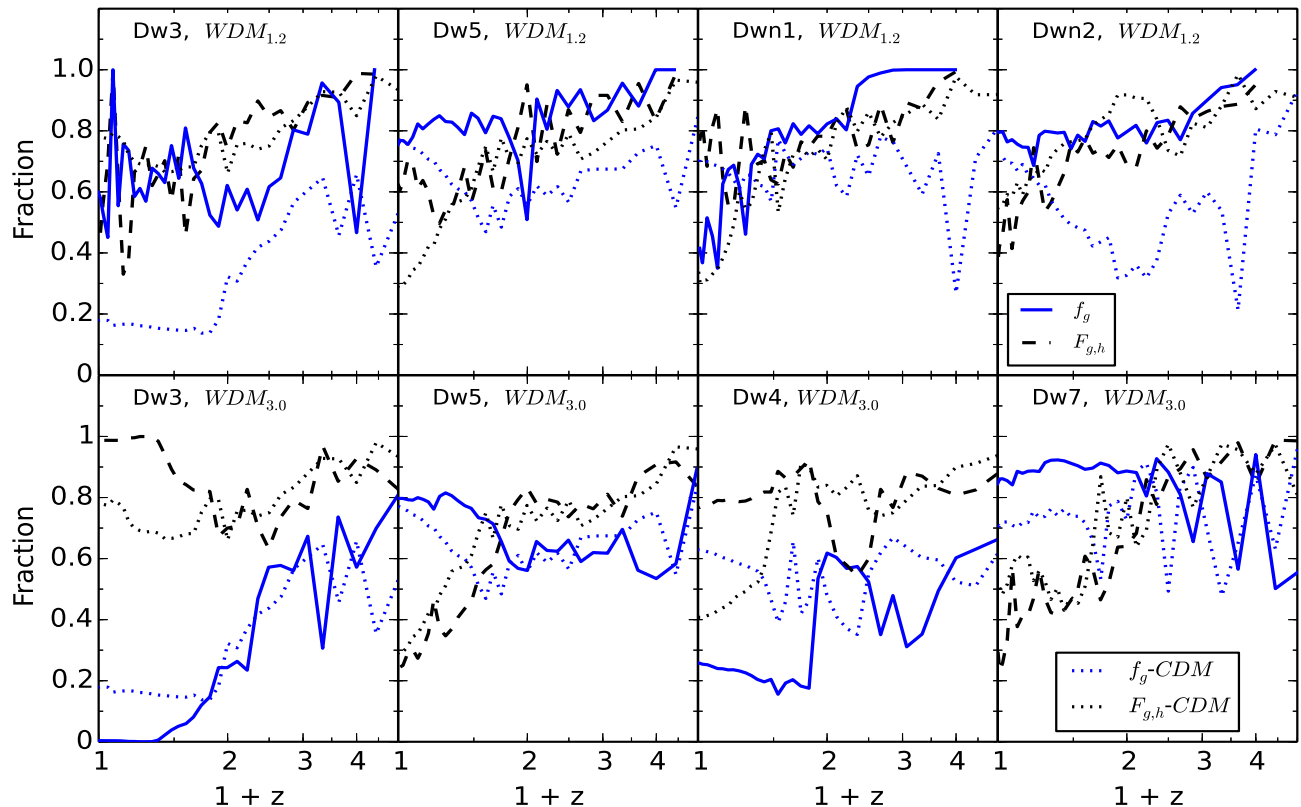}
\caption[fig6]{
Evolution of the galaxy gas mass fraction, \fg, for the different WDM runs (blue solid lines). 
The black dashed line refers to \Fgh, the ratio between the gas mass in the halo (the gas in the
spherical shell of radii 0.1\rh\ and 1\rh) and the total gas mass within \rh. 
 With blue and black dotted lines,  \fg\ and \Fgh, respectively,  
we show the corresponding quantities for the CDM case. The \wdma\ runs, plotted in the upper panels, 
present evolutions of \fg\ and \Fgh\ that differ significantly from those of their 
CDM counterparts, while, on ther other hand, for the \wdmb\ runs, plotted in the lower panels, 
the differences are less notorious.
}
\label{fig6}
\end{figure*}
%========================

The blue solid and dashed lines in Fig. \ref{fig5} show the stellar MAHs of the WDM and CDM 
simulated dwarf galaxies, respectively.  In G+2014, it was shown that the stellar MAHs of dwarfs
follow roughly their halo MAHs in the CDM simulations; i.e., the \ms-to-\mh\ ratio is roughly constant,
at least up to $z\sim 2$ (with variations of $0.1-0.4$ dex).  
Here, for the \wdma\ dwarfs we find that from $z\gtrsim 2$ to $\sim 1$ the \ms-to-\mh\
ratio significantly increases; from $z\sim 1$ to 0, this ratio continues increasing (except in run Dw5)
but only slightly. The early fast increase occurs because at earlier epochs the \wdma\ baryonic 
galaxies are in their active growth phase. Thus, the stellar mass assembly of the dwarfs
at the filtering scale shows a decoupling from the assembly of their halos, unlike 
what happens with the dwarfs in the CDM case. 
At $z=0$, the \ms-to-\mh\ ratios of all \wdma\ galaxies are $2-4$ times lower
than their CDM counterparts except for Dw3 in which case they are similar. 
As expected, the differences are much smaller when compared to the \wdmb\ runs; 
at $z = 0$, the maximum difference, which amounts a factor of 1.6, is found for the dwarf Dw7.
In general, the \ms-to-\mh\ ratio remains close to the CDM one at all redshifts.
Thus,  as the halo mass gets closer to the filtering mass, the galaxies formed inside them
end up with lower \ms-to-\mh\ ratios. This result strictly holds for the scales 
and neutrino masses (1.2 and 3.0 keV) studied here.

Our \wdma\ dwarf galaxies
assemble their stellar masses with a significant delay with respect to their CDM counterparts. 
For example, the assembly of half of the present-day \ms\ for dwarf Dw5 happens 1.3 Gyr 
later in the \wdma\ cosmology. For the dwarfs formed in halos much larger than \Mhm\  (runs \wdmb), 
the stellar MAHs are close to those of the CDM counterparts,
with minimum differences in the half-mass assembly epochs. 

The galaxy baryon (stars + gas) MAHs, \mb($z$), of the simulated galaxies follow moderately the halo MAHs,
with some intermittence, both in the WDM and CDM simulations. However, the \wdma\ runs show 
more intermittent histories than the CDM ones, due to more extended periods of gas infall/outflow 
onto/from the halos. 
This is likely because the \wdma\ halos accrete baryons in a more regular way than the CDM halos
and because they blowout the gas more efficiently due their lower concentrations and \vmax.
The baryon-to-halo mass ratios, \mb-to-\mh, of the former are slightly lower than those of 
the latter at all epochs, except for run Dw3 at $z \lesssim 0.7$ . In particular, 
at $z=0$, the \mb-to-\mh\ ratios of the \wdma\ galaxies are $1.1-3.5$ times lower 
than their CDM counterparts.
\begin{figure*}
\vspace{12.0cm}
\includegraphics{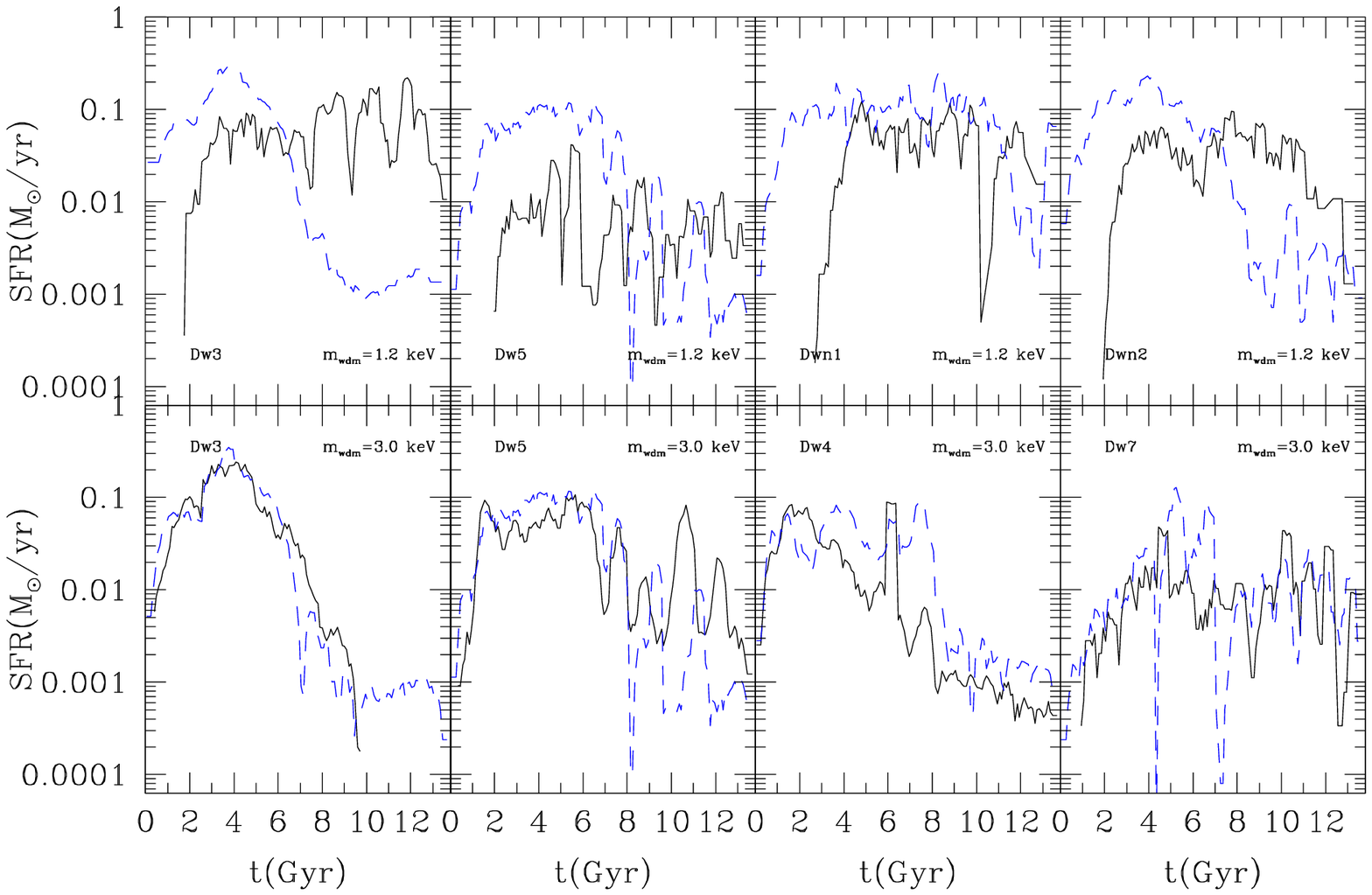}
\caption{
Archaeological SF rate histories for the \wdma\ (upper panels) and \wdmb\ (lower panels) runs 
presented in this paper (black solid lines). In each panel, the corresponding CDM 
SF histories are also shown (dashed blue lines). 
In the x-axis runs the cosmic time. The histories were smoothed with a top-hat filter of 
500 Myr width; see text for details about how the SF histories were calculated.}
\label{fig7}
\end{figure*}
%========================

%=========================================================
\subsection{Gas fraction and star formation histories}

In Fig. \ref{fig6}, we plot the change with redshift of the {\it galaxy} gas fractions (\fg=\mg/\mb, 
solid lines) for the \wdma\  and \wdmb\ runs, upper and lower panels, respectively.  
The dashed lines show the change with redshift of the fraction of gas outside the galaxy 
but within the {\it halo}, the``halo'' gas fraction \Fgh. This is defined as the 
ratio of the gas mass contained in the spherical shell of radii between 0.1\rh\ and \rh\ to the 
gas mass in the whole halo.  As in the case of the CDM simulations (see Fig. 6 in G+2014), 
the two gas fractions oscillate, and in periods where \fg\ decreases (increases)
typically \Fgh\ increases (decreases). This is mainly due to the interplay among gas 
accretion onto the galaxy, SF and SN-driven outflows from the galaxy. It seems that this 
interplay is stronger in the CDM simulations than in the systems with
mass around \Mhm\ (\wdma). The \wdmb\ runs show a behavior roughly close to their CDM counterparts.

The galaxy gas fractions in the \wdma\ runs are, at all epochs, higher than those found in
their CDM runs counterparts, probably as a consequence of the later galaxy assembly of 
the former runs. The values of \fg\ at $z=0$ are shown in column eighth of Table 1. We see that
the \fg\ values of the WDM dwarfs formed in halos of scales around \Mhm\ are relatively high.

In Fig. \ref{fig7} we plot the "archeological" SF histories of our WDM runs (solid lines)
compared to their CDM counterparts (blue dashed lines). 
For illustration purposes, since the SF histories are strongly intermittent, they 
were smoothed with a top-hat filter of 500 Myr width. The original histories
were built with 0.1 Gyr bins. This is computed, for any given time $t$ [Gyr], 
identifying all galaxy stellar particles at $z=0$ born within the time interval [$t-0.1,$t] Gyr; the SFR
at this time is then simply the mass of these particles divided by 0.1 Gyr.

As expected from the later halo assembly, the SF in  the \wdma\  systems of masses around \Mhm\ 
starts later than in the CDM counterparts. Besides, the former present {\it more sustained SF histories at later
epochs than the CDM dwarfs}, for which the SFR tends to fall in the last Gyrs. This implies that the 
specific SFRs (SFR/\ms) of the WDM galaxies of scales around \Mhm\ tend to be higher at late epochs than
those of the CDM counterparts.  
The fact that the \wdma\ galaxies assemble later and have lower central gas surface densities than the CDM ones
likely explains their less efficient but more sustained SF histories. 
As in the CDM runs (see a discussion in G+2014), the SF histories in the WDM runs are also episodic.
For those systems around the filtering mass \Mhm, the SF is sometimes even more bursty than their
CDM counterparts. 

To highlight the differences in the stellar populations between the dwarfs in the \wdma\ runs
and their CDM counterparts, in Fig. \ref{fig8} we plot the cumulative 
(archaeological) SF histories. Solid lines are for the \wdma\  and \wdmb\ dwarfs, left
and right panel, respectively, while dashed lines  in both panels are for
the CDM counterparts. One clearly sees that the stellar
populations of present-day \wdma\ dwarfs are formed on average significantly later 
than those of the CDM dwarfs, with $20\%$ of their stars being formed in the last $\sim 4$ Gyr; 
in contrast, the CDM dwarfs, have formed already
80\% of their stars between $\sim 7.5$ and 9 Gyr ago (Dwn1 reaches this fraction later, 
$\sim 4.6$ Gyr ago). The galaxy Dw3 is the one that most differs in its SF history (black lines), 
when compared \wdma\ with CDM, and Dwn1 is the one that shows the most similar history (blue lines).
 Interestingly, it is the galaxy Dw3 the one that differs less when compared \wdmb\ with CDM. In
general, the differences between the \wdmb\ and CDM galaxies (right panel), as expected, are lower
than those found when the \wdmb\ and CDM dwarfs are compared.

We have calculated also the mass-weighted ``archeological'' ages of all runs and report them in 
the last column of Table 1. This age is the result of multiplying the age of each galaxy
stellar particle at $z = 0$ by its mass fraction contribution (the particle mass divided by \ms), 
and summing these terms for all the particles.
{\it The dwarfs of scales around \Mhm\ are between $\sim 1.4$ and 4.8 Gyrs younger than their
CDM counterparts.} The largest difference is for the dwarf Dw3
and the smallest for Dwn1 (see above). The mass-weighted ages of the \wdmb\ dwarfs are similar
or slightly smaller (by $\sim 1$ Gyr) than the CDM counterparts.
In summary, {\it central dwarf galaxies in the WDM scenario are expected to have 
younger stellar populations on average than their CDM counterparts,} the younger the 
closer their halo masses are to the filtering scale.

%=========================================================
\section{Summary and Discussion}
\label{discussion}
%=========================================================

We have presented the first N-body + Hydrodynamics (zoom-in) simulations of
galaxies formed in distinct WDM halos with  masses at present-day 
close to the half-mode filtered mass \Mhm\ corresponding to a thermal WDM particle mass of 
\mwdm=1.2 keV. Halo masses are around $3 \times 10^{10}\ \msun$. Galaxies 
formed in WDM halos 20--30 times more massive than \Mhm\ were also simulated 
(runs \wdmb,  for which \mwdm=3.0 keV). In a WDM cosmology, 
the halos of masses around \Mhm\ are close to the peak of the halo mass function; at 
masses a factor of $\sim 2-3$ lower, the halo mass function declines sharply 
due to the damping of the initial power spectrum of fluctuations
\citep[e.g.,][]{Schneider+2013,Angulo+2013}. Most of structures $\sim 3$ times smaller
than \Mhm\ at $z=0$ already do not appear to be virialized spherical overdensities
(halos) and they did not assemble hierarchically.

\begin{figure}
\includegraphics[width=8.5cm]{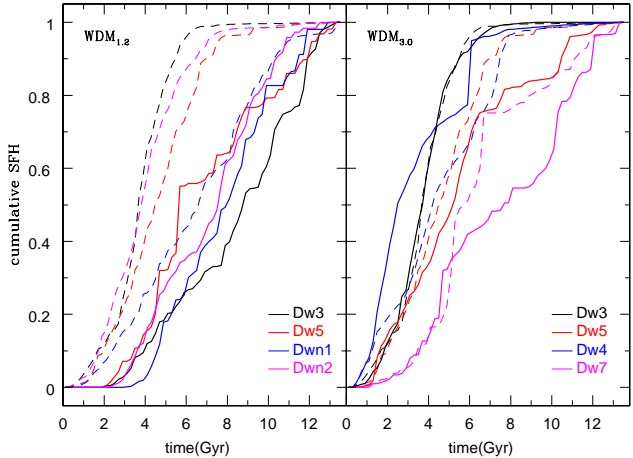}
\caption{Cumulative SF histories of the \wdma\  and \wdmb\ runs (solid lines), left and right panel, respectively,
along with their CDM counterparts (dashed lines  in both panels ). 
The different dwarfs are identified with different colors. 
The \wdma\ runs clearly form most of their present-day stars much later than their CDM counterparts.  As
expected, the differences between the \wdmb\ and CDM runs (right panel) are lower than those found in the left panel,
although there are runs and periods in which these differences are significant.
}
\label{fig8} 
\end{figure}

Our results show that  the \wdma~ galaxies have
disk-like structures and circular velocity profiles that gently increase and then flattens.
These dwarfs are quite different in several aspects from their CDM
counterparts, which assembled hierarchically. The galaxies formed in halos 
$20-30$ times \Mhm\  (runs \wdmb), instead, are very similar in properties
and evolution to their CDM counterparts, in agreement with the results of
\citet{Herpich+2014}.
Therefore, {\it the properties and evolution of WDM galaxies differ more
from those of the CDM galaxies as the mass get closer to the filtering scale.}
In summary,  our \wdma\ dwarf galaxies that formed in halos with a mass 
around \Mhm\ differ from their CDM counterparts in that:
\begin{enumerate}
\item they assemble their stellar masses later (Fig. \ref{fig5}), with
archaeological SF histories shifted to younger stellar populations 
(Fig. \ref{fig8}; on average, the WDM dwarfs have
 mass-weighted ages 1.4--4.8 Gyr younger than the CDM ones, see also
\citealp{Governato+2014});
\item their \vmax\ values are 20--60\% lower;
\item they have significantly lower central stellar SDs and 
larger \re\ values, by factors of 1.3--3 (Fig. \ref{fig4});
\item their $V_c(r)$ profiles are shallower, being this mainly
because the baryonic (stars + gas) components are shallower (Fig. \ref{fig3});
\item on average, they have higher gas fractions and lower
 stellar masses (and thus lower \ms-to-\mh\ ratios).
 \end{enumerate}
 
As stated above, the reported differences were found in
galaxies formed in halos with the particular value for the
filtering scale of $\Mhm=2\times 10^{10}$ \msunh\
($\mwdm= 1.2\ \kev$). Can these results be
generalized to other scales (or WDM particle masses)?
We could argue that at least qualitatively our results might be extended to other masses
because part of the these differences, lower Vmax, \ms-to-\mh\ ratios, etc., are systematically
due to the lower concentrations and delay in the formation of the WDM halos.
Nevertheless, in general, the results of our hydrodynamic simulations should not be
rescaled with respect to the filtering mass because the astrophysical processes such as cooling,
feedback, etc., affect significantly the evolution of the galaxy-halo systems, specially in small
halos (higher WDM particle masses).
 
%==========================
%==========================

\subsection{WDM galaxy formation}

If the Ly-$\alpha$ power spectrum constrains WDM models to be made of
relic particles with masses above $\approx 3$ keV \citep{Viel+2013}, then the 
corresponding filtering scale at $z=0$ should be $\lesssim 1.5\times 10^9$ \msun.  
In the CDM cosmology, the distinct halos of $1.5\times 10^9$ \msun~
have \vmax\ values of $\sim 20$ km/s and their stellar masses are expected
to be $\lesssim 10^7-10^6$ \msun\ (\ms-to-\mh\ ratios $\lesssim 7-0.7\times
10^{-3}$), depending on the used subgrid physics
 \citep[e.g.,][]{Sawala+2010,Munshi+2013,Cloet-Osselaer+2014,Sawala+2014a,Sawala+2014b,Governato+2014}.

 If we now extend our results found in this work; that is, the 
differences found between the \wdma\ and
CDM simulations, to the hypothetical dwarfs formed
in halos at the filtering scale of $1.5\times 10^9$ \msun\ (\mwdm=3 keV), then the corresponding $\vmax$ and 
$\ms/\mh$ ratio would be around 12 km/s and  $0.005$, respectively. On the other hand, 
we would expect these WDM dwarfs to have higher gas fractions, lower central stellar 
SDs and later SF histories than the corresponding CDM ones. 
 As mentioned above, the results of the hydrodynamic simulations may or may not be
generealized to other filtering scales, so this extrapolation of our results 
should be considered just as a 
qualitative statement. 

Unfortunately, current observations of field dwarf galaxies of masses and circular velocities as
small as those corresponding to halos of masses $\sim 10^9$ \msun\ are so 
limited that they can not be used to distinguish 
between the CDM and the WDM with $\mwdm\approx 3$ keV cosmogony.
The few and uncertain observations
of central (field) very small dwarfs point out that practically all of them
are star-forming and gaseous rich galaxies
\citep[e.g.,][]{Geha+2006, Geha+2012}, with late SF histories
\citep[e.g.,][]{Weisz+2014, Cole+2014}, and with \vmax\ values
for a given \ms\ smaller than those inferred or simulated in the CDM
scenario \citep[e.g,][]{Ferrero+2012,Rodriguez-Puebla+2013}, thus favoring 
the WDM scenario. However, as several authors have shown, these potential 
disagreements in the CDM scenario, in particular those related to the too-high 
\vmax\ and \ms/\mh\ values, could be solved also by plausible changes/improvements 
in the subgrid physics; for example, by introducing a
metallicity-dependent H$_2$ molecule formation process
\citep[][]{Kuhlen+2012,Christensen+2012} or by
introducing preventive/early mechanisms of feedback besides of increasing 
the strength of the ejective SN-driven feedback
\citep{Hopkins+2012,Hopkins+2013,Munshi+2013,Trujillo-Gomez+2013,Stinson+2013,Agertz+2013}.

Along this venue, \citet[][see also a recent review by Brooks 2014)]{Governato+2014}  
argue that the effects of the SF-driven feedback overcome those of the
initial power spectrum regarding the inner dark matter and stellar
mass distributions. This conclusion is based on only one zoom-in simulated dwarf in both
CDM and WDM cosmologies. For the latter, the filtering corresponds to a relic particle 
of mass 2 keV, which means that $\Mhm=5.7\times 10^9$ \msun\ 
(see Fig. \ref{fig1}). The present-day virial mass of their dwarf is
$\approx 1.4\times 10^{10}$ \msun\ (after correcting by a factor of 1.23
as one goes from $M_{200}$ to \mh); that is, this system is $\approx 2$ times larger
than the filtering mass. For this particular object, the dark-matter only simulations in 
the CDM and WDM cosmologies show that their $V_c(r)$ profiles are actually
not too different from each other (see their Fig. 8). 

 At the level of dark-matter only simulations, the four \wdma\ halos (to be presented elsewhere, 
Avila-Reese et al. in preparation) show different \vmax\ values and $V_c(r)$ profiles as compared with 
the corresponding CDM halos. There are also significant differences regarding the assembly histories. 
Thus, the effects of the damping of the power spectrum seem to have 
significant effects on the structures close to \Mhm\ already in pure N-body simulations. 
WDM halos at the cutoff of the power spectrum are certainly different than the CDM ones and, when baryons are
included in the simulations, the initial conditions could leave an imprint in 
the respective galaxies. 
Note that the astrophysical effects also affect the dark matter halo properties
so that predictions based on dark-matter only results that are then compared
to observations should be taken with care \citep[for example, when comparing
the WDM halo velocity function to the observed galaxy velocity function][]{Zavala+2009,Klypin+2014,Papastergis+2014}

Future observational studies of central (field) dwarf galaxies will be
crucial for constraining the nature of dark matter.
In addition to the inner dynamics, we have also found important differences
between CDM and WDM dwarfs in their SF histories, stellar SD profiles 
(specially in the central regions), and gas fractions.

It should be said that resolution issues are likely affecting our results
regarding the earliest stages ($z>3$) of the WDM galaxy assembly, where 
virial masses get closer to the scale of artificial fragmentation of filaments and 
to the free-streaming scale. Very high-resolution simulations,
including baryons, suggest that structures around the free-streaming scale
are smooth and dense filaments able to capture gas that can
cool efficiently and form stars \citep{Gao+2007, Gao+2014}.
Thus, the smallest (earliest) baryonic structures in a WDM cosmology are
expected to be filament-like; certainly, the formation of stars (the first ones) 
in this environment is different from that in a virialized halo
\citep[see][]{Gao+2007}.  SF may efficiently proceed in these filaments
before they disappear into the more familiar halo-like structures, so that a non-negligible fraction
of stars in the $z=0$ galaxy may have formed early in these 
filaments. Hence, our result that the fraction
of stars formed during the first 2--4 Gyr in the \wdma\ runs is negligible
(Fig. \ref{fig8}) could be an underestimation 
due to our inability to adequately resolve and follow the physics of
the gas in the first smallest filaments (their masses should be of the order of
the corresponding free-streaming mass, $\sim 2\times 10^6$ \msunh).

%==========================
\begin{figure}
\includegraphics[width=8.0cm]{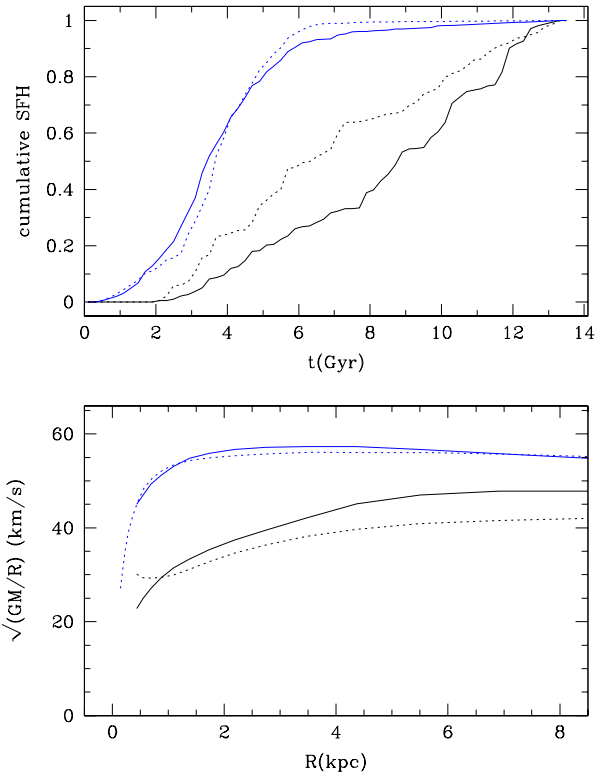}
\caption{A comparison of the CDM  and \wdma\ runs, blue and black lines, respectively,
of dwarf Dw3 with aggressive (dotted lines) and soft 
(solid lines) refinements is presented. In the upper panel we show the comparison regarding
the cumulative SF history while in the lower panel the comparison is with respect to 
the circular velocity (see discussion in the appendix).}
\label{fig9} 
\end{figure}
%==========================

We end the discussion by asking whether our simulated dwarfs with \mwdm=1.2 keV are in
agreement with observations. In \citet{Avila-Reese+2011} and in G+2014 we studied the
properties and evolution of CDM low-mass galaxies, some of which (Dw3, Dw4, Dw5 y Dw7) were also studied
here. These CDM galaxies with total masses around $1-5\times10^{10}$ \msunh\
are relatively realistic in structural and dynamical properties; however, they have lower
specific SF rates, too high \ms-to-\mh\ ratios and lower gas fractions than 
the observed ones, showing that they form most of their stars too early.
The \wdma\ dwarfs simulated here {\it with the same subgrid physics} have
delayed SF histories, form less stars, and have more gas than their CDM counterparts.
 However, when compared to observed galaxies of similar stellar masses, they are too extended.
In any case, a WDM model with \mwdm=1.2 keV seems to be in conflict with the last
Ly-$\alpha$ forest constraints \citep{Viel+2013}.

\section*{Acknowledgements}            
We are grateful to the Referee for his/her constructive comments. VA and AG  acknowledge 
CONACyT grant (Ciencia B\'asica) 167332. AG acknowledges a PhD fellowship provided by DGEP-UNAM

\appendix

As mentioned in Section 3.1, the CDM dwarfs were run with the same refinement setting used in G+2014.
However, to try to reduce the probable appearance and growth of spurious fragments, a less aggressive 
refinement was used in the WDM runs. 
This means that WDM dwarfs are slightly less resolved than their CDM counterparts; 
that is, the halo/galaxy ends up with less resolution elements. To show that no significant differences appear
when this setting (resolution) is used, we run some of our CDM simulations with this soft refinement. In Fig.
\ref{fig9}, we compare the cumulative SFH (upper panel) and circular velocity 
(lower panel) of the CDM run of Dw3 using 
the aggressive refinement (blue dotted lines) with the corresponding quantities of the less aggressive setting
(blue solid lines). The difference between the two runs in \vmax\ amounts only 
to $\sim$ 2\%, while in the cumulative SFH the major differences are seen during the stage of active star
formation, in the first $\sim$ 5 Gyr. On the other hand, 
the late star formation seems to be somewhat
sensitive to the details of the used resolution. 
Yet, we certainly believe this should not be much of concern because 
the amount of mass in stars formed in this relatively quiet star forming phase is small; in general,
the stellar mass \ms\ in the less resolved run is only 17\% higher than the corresponding one with the 
aggressive refinement setting.

In Fig. \ref{fig9}, we also plot the \wdma\ runs for the same Dw3 galaxy. The use of an aggressive 
refinement (black dotted lines), produces older stellar populations than in the less aggressive 
refinement employed in the paper (black solid lines). The refinement setting clearly affects 
more the \wdma\ run than the CDM one. This is very likely related to the greater number of
spurious fragments form at early times in the more aggressive setting, which are incorporated later to 
the halo/galaxy. The differences found in the $V_c$ profile are also greater
in the refinement setting in the \wdma\ run than in the CDM one. In particular, we see that 
the \wdma\ simulation with the more aggressive refinement (black dotted line) forms a centrally 
mass concentration (inner peak) composed of mass in stars and dark matter.
This is consistent with the higher early SFR seen in the top panel. In any case, the differences 
in the SF history and $V_c$ profile 
between the \wdma\ and CDM Dw3 run {\it remain} qualitatively the same in our simulations 
regardless of the refinement setting.

\end{document}